\documentclass[aps,prd,twocolumn,groupedaddress,amsfonts,amssymb,amsmath,showpacs,showkeys]{revtex4-1}

\usepackage{graphicx}

\usepackage{psfrag}

\begin{document}


\title{Wide binaries as a critical test of Classical Gravity}

\author{X. Hernandez$^{1}$}
\email[Email address: ]{xavier@astroscu.unam.mx}
\author{M. A. Jim\'enez$^{1}$}
\email[Email address: ]{mjimenez@astro.unam.mx}
\author{C. Allen$^{1}$}
\email[Email address: ]{chris@astroscu.unam.mx}
\affiliation{$^1$Instituto de Astronom\'{\i}a, Universidad Nacional
                 Aut\'onoma de M\'exico, AP 70-264, Distrito Federal 04510,
	         M\'exico}

\date{\today}

\begin{abstract}


Modified gravity scenarios where a change of regime appears at acceleration scales $a<a_{0}$ have been proposed. 
Since for $1 M_{\odot}$ systems the acceleration drops below 
$a_{0}$ at scales of around 7000 AU, a statistical survey of wide binaries with relative velocities and separations reaching 
 $10^{4}$ AU and beyond should prove useful to the above debate. We apply the proposed test to the best currently 
available data.
Results show a constant upper limit to the relative velocities in wide binaries which is 
independent of separation for over three orders of magnitude, in analogy with galactic flat rotation curves in the 
same $a<a_{0}$ acceleration regime. Our results are suggestive of a breakdown of Kepler's third law 
beyond $a \approx a_{0}$ scales, in accordance with generic predictions of modified gravity theories designed not to require 
any dark matter at galactic scales and beyond.
\end{abstract}


\maketitle

\section{Introduction}
\label{introduction}

Over the past few years the dominant explanation for the large mass to light ratios inferred for galactic and meta-galactic 
systems, that these are embedded within massive dark matter halos, has begun to be challenged. Direct detection of the dark 
matter particles, in spite of decades of extensive and dedicated searches, remains lacking. This has led some to interpret the 
velocity dispersion measurements of stars in the local dSph galaxies (e.g. \citep{2}, \citep{14}), the extended and 
flat rotation curves of spiral galaxies (e.g. \citep{26}, \citep{21}), the large velocity dispersions of galaxies 
in clusters (e.g. \citep{23}), stellar dynamics in elliptical galaxies (e.g. \citep{34}),
the gravitational lensing due to massive galaxies (e.g. \citep{33}, \citep{8}), 
and even the cosmologically inferred matter content for the universe through CMB and structure formation physics
(e.g. \citep{30}, \citep{13}, \citep{24}), not as indirect evidence for the existence of a dominant dark matter
component, but as direct evidence for the failure of the current Newtonian and General Relativistic theories of gravity, in the 
large scale or low acceleration regimes relevant for the above situations.

Numerous alternative theories of gravity have recently appeared (e.g. TeVeS of \citep{3}, and variations;  
\citep{5}, \citep{31}, F(R) theories e.g. \citep{29}, \citep{6}, \citep{35}, conformal 
gravity theories e.g. \citep{19}), mostly grounded on geometrical extensions to General Relativity, and
leading to laws of gravity which in the large scale or low acceleration regime, mimic the MOdified Newtonian Dynamics (MOND) 
fitting formulas. Similarly, \citep{22} have explored MOND not as a modification to Newton's second law, but as a 
modified gravitational force law in the Newtonian regime, finding a good agreement with observed dynamics across galactic scales 
without requiring dark matter. In fact, recently \citep{4} have constructed an $f(R)$ extension to general relativity 
which in the low velocity limit converges to the above approach.

Whilst Classical Gravity augmented by the dark matter hypothesis provides a coherent and unified interpretation from galactic to 
cosmological scales (with the inclusion of dark energy), the very profusion of modified gravity theories, mostly tested in very 
localised situations, points to the lack of any definitive theoretical contender to Classical Gravity. Nonetheless, a generic 
feature of all of the modified gravity schemes mentioned above is the appearance of an acceleration scale, $a_{0}$, above which 
classical gravity is recovered, and below which the dark matter mimicking regime appears. This last feature results in a general 
prediction; all systems where $a>>a_{0}$ should appear as devoid of dark matter, and all 
systems where $a<<a_{0}$ should appear as dark matter dominated, when interpreted under classical gravity.  It is interesting 
that no $a>>a_{0}$ system has ever been detected where dark matter needs to be invoked, in accordance with the former condition. 
On the other hand, the latter condition furnishes a testable prediction, in relation to the orbits of wide binaries. For test 
particles in orbit around a $1 M_{\odot}$ star, in circular orbits of radius $s$, the acceleration is expected to drop below 
$a_{0}\approx 1.2 \times 10 ^{-10} m/s^{2}$ for $s>$7000 AU$=3.4\times10^{-2} pc$. The above provides a test for the dark matter/
modified theories of gravity debate; the relative velocities of components of binary stars with large physical separations should 
deviate from Kepler's third law under the latter interpretation.

More specifically, seen as an equivalent Newtonian force law, beyond $s \approx $7000 AU the gravitational force 
should gradually switch from the classical form of $F_{N}=GM/s^{2}$ to $F_{MG}=(G M a_{0})^{1/2}/s$, and hence the 
orbital velocity, $V^{2}/s =F$, should no longer decrease with separation, but settle at a constant value, dependent 
only on the total mass of the system through $V=(G M a_{0})^{1/4}$. That is, under modified gravity theories, 
binary stars with physical separations beyond around 7000 AU should exhibit ``flat rotation curves'' and a 
``Tully-Fisher relation'', as galactic systems in the same acceleration regime do.

An interesting precedent in this sense is given by the recent results of \citep{27} and  \citep{28}  
who find evidence for a transition in the dynamics of stars in the outer regions of a series of Galactic globular
clusters. These authors report a flattening of the velocity dispersion profile in globular clusters, outwards of
the radius where accelerations fall below the $a_{0}$ threshold, in accordance with generic predictions of modified
gravity schemes. The interpretation under Newtonian dynamics explains the observed flattening as due to tidal heating
by the Milky Way, e.g \citep{18}, but the matter is still being debated.   We also note the recent results of \citep{17} 
who point out various discrepancies between standard $\Lambda CDM$ predictions and structural and dynamical 
properties of the local group,  and suggest solutions to these in the context of modified gravity theories.

In this paper we propose that wide binary orbits may be used to test Newtonian gravity in the low acceleration regime.  We apply 
this test to the binaries of two very recent catalogues containing relative velocities and separations of wide binaries.
The two catalogues are entirely independent in their approaches. The first one, \citep{29} uses data from the {\it 
Hipparcos} satellite to yield a moderate number of systems (280) relatively devoid of false positives (10\%), with a high average 
signal to noise ratio for the relative velocities of the binaries ($\sim 2$).  The second, \citep{11} identifies 1,250 
wide binaries from the Sloan Digital Sky Survey (SDSS) data base data release 7, which, compounded with a detailed galactic 
stellar distribution model, results in pairs with a very low probability of chance alignment ($<2\%$), albeit with a low average 
signal to noise ratio in their relative velocities ($\sim 0.5$).

The paper is organised as follows: section (2) briefly gives the expectations for the distribution of relative velocities as a 
function of separation for wide binaries, under both Newtonian gravity and generically for modified theories of gravity. In 
section (3) we show the results of applying the test to the \citep{29} {\it Hipparcos} catalogue, and to the 
independent \citep{11} SDSS data. Our conclusions are summarised in section (4).

\section{Expected relative velocity distributions for wide binaries}

Since orbital periods for $1 M_{\odot}$ binaries with separations in the tens of AU range already extend into the centuries, 
there is no hope of testing the prediction we are interested in through direct orbital mapping. Fortunately, modern relative 
proper motion studies do reach binary separations upwards of $10^{4}$ AU, e.g. \citep{1}, \citep{7}, \citep{11}. 
The Newtonian prediction for the relative velocities of the two components of binaries having circular 
orbits, when plotted against the binary physical separation, $s$, is for a scaling of $\Delta V \propto s^{-1/2}$, essentially 
following Kepler's third law, provided the range of masses involved were narrow. 

In a relative proper motion sample however, only two components of the relative velocity appear, as velocity along the line of 
sight to the binary leads to no proper motion. Thus, orbital projection plays a part, with systems having orbital planes along 
the line of sight sometimes appearing as having no relative proper motions. A further effect comes from any degree of orbital 
ellipticity present; it is hence clear that the trend for $\Delta V \propto s^{-1/2}$ described above, will only provide an upper 
limit to the distribution of projected $\Delta V$ vs. $s$ expected in any real observed sample, even if only a narrow range of 
masses is included. One should expect a range of measured values of projected $\Delta V$ at a fixed observed projected $s$, all 
extending below the Newtonian limit, which for equal mass binaries in circular orbits gives:

\begin{equation}
\Delta V_{N} =2 \left( \frac{G M}{s} \right)^{1/2}.
\end{equation}

The problem is complicated further by the dynamical evolution of any population of binaries in the Galactic environment. Over 
time, the orbital parameters of wide binaries will evolve due to the effects of Galactic tidal forces. Also, dynamical encounters 
with other stars in the field will modify the range of separations and relative velocities, specially in the case of wide 
binaries. To first order, one would expect little evolution for binaries tighter than the tidal limit of the problem, and the 
eventual dissolution of wider systems. 

A very detailed study of all these points has recently appeared, \citep{16}. These authors numerically follow 
populations of 50,000 $1 M_{\odot}$ binaries in the Galactic environment, accounting for the evolution of the orbital parameters 
of each due to the cumulative effects of the Galactic tidal field at the Solar radius.  Also, the effects of close and long range 
encounters with other stars in the field are carefully included, to yield a present day distribution of separations and relative 
velocities for an extensive population of wide binaries, under Newtonian Gravity. Interestingly, one of the main findings is that 
although little evolution occurs for separations below the effective tidal radius of the problem, calculated to be of 1.7 pc, the 
situation for grater separations is much more complex than the simple disappearance of such pairs.

It is found that when many wide binaries cross their Jacobi radius, the two components remain fairly close by in both coordinate
and velocity space, drifting in the Galactic potential along very similar orbits. This means that in any real wide binary search 
a number of wide pairs with separations larger than their Jacobi radii  will appear. Finally, \citep{16} obtain the 
RMS one-dimensional relative velocity difference, $\Delta V_{1D}$, projected along an arbitrary line of sight, for the entire 
populations of binaries dynamically evolved over 10 Gyr to today, for a distribution of initial ellipticities, as plotted against 
the projected separation on the sky for each pair. The expected Keplerian fall of $\Delta V_{1D} \propto s^{-1/2}$ for 
separations below 1.7 pc is obtained, followed by a slight rise in $\Delta V_{1D}$ as wide systems cross the Jacobi radius 
threshold. $\Delta V_{1D}$ then settles at RMS values of $ \approx 0.1 km/s$.

These authors also explore variation in the initial (realistic) distribution of semi-major axes, and the formation history 
of the binaries, finding slight differences in the results, which however are quite robust to all the variations in the 
parameters explored, in the regime we are interested of present day projected separations larger than $log (s/pc)>-2$, 
above very small variations of less than  0.14 in the logarithm. This represents the best currently available estimate of how 
relative velocities should scale with projected separations for binary stars (both bound and in the process of dissolving in 
the Galactic tides) under Newtonian gravity.

The testable quantitative prediction of Classical Gravity for the distribution of data in a plot of projected $\Delta V$ vs.
projected $s$ is clear: one should find a spread of points extending below the limit defined by eq.(1), with an RMS value for 
$\Delta V_{1D}$ given by the results of \citep{16}, their figure (7).

The recent proliferation of modified gravity models however, implies the absence of a definitive alternative to Classical 
Gravity. Further, in many cases, the complex formulations put forth do not lend themselves to straightforward manipulations from 
which detailed predictions might be extracted for varied applications distinct from the particular problems under which such 
models are presented. We shall therefore not attempt to test any particular modified gravity theory, but shall only consider the  
generic predictions such theories make for "flat rotation curves" in the $a<a_{0}$ regime. That is, the predictions of 
modified gravity schemes will only be considered qualitatively and generically, to first order, in terms of the upper envelope of 
observed distributions in projected $\Delta V$ vs. $s$ plots to appear flat. For circular orbits one expects:

\begin{equation}
\Delta V_{MG} = C (G M a_{0})^{1/4} ,
\end{equation}

\noindent where $C$ is a model-dependent constant expected to be of order unity. A further correction upwards due to departures 
from circularity, which at this point must be thought of as dependent on the details of the particular modified gravity scheme 
one might pick, should also be included. This correction will tend to give even larger values of $\Delta V_{MG}$. We note that
in the particular case of MOND, the external field effect, the fact that the overall potential of the Galaxy at the solar 
neighbourhood globally puts local binaries close to the $a=a_{0}$ threshold, would imply only slight corrections om Newtonian
predictions.

We see that all we need is a large sample of relative proper motion and binary separation measurements to   
test the Newtonian prediction for the RMS values of the 1 dimensional relative velocities of \citep{16},
and to contrast the $\Delta V_{N} \propto s^{-1/2}$ and the $\Delta V_{MG} =cte.$ predictions for the upper envelope
of the $\Delta V$ vs. $s$ distributions. It is important to have a sample as free of chance alignments as possible, as the 
inclusion of non-physical stellar pairs would blur the test, potentially making a conclusion suspect. Also, it is desirable to 
limit the range of masses of the stars involved, as a spread in mass will also blur any trends expected for the upper limit of 
the $\Delta V$ distributions, although not terribly so, given the small powers to which mass appears in both predictions.

To end this section we briefly recall the first order tidal limit calculation of 

\begin{equation}
 \left. \frac{d F_{ext}(R)}{dR} \right|_{R_{0}}  \Delta r = \frac{G M_{s}^{2}}{(\Delta r)^{2}}
\end{equation}

\noindent which leads to the tidal density stability condition of $\rho_{s}>\overline{\rho}$ for the density of a satellite 
of extent $\Delta r$ and mass $M_{s}$ orbiting at a distance $R_{0}$ from the centre of a spherical mass distribution $M(R)$ 
having an average matter density $\overline{\rho}$ internal to $R_{0}$ resulting in a gravitational force $F_{ext}(R)$, 
under the assumption $\Delta r <<R$. The equivalent calculation under the force law given by the $a<<a_{0}$ limit of 
$F_{MG}=(G M a_{0})^{1/2}/R$ is given by:

\begin{equation}
\frac{(G M(R) a_{0})^{1/2}}{R^{2}}\Delta r  = \frac{(G M_{s} a_{0})^{1/2}}{\Delta r},
\end{equation}

\noindent leading to:

\begin{equation}
\rho_{s}> \left(\frac{\Delta r} {R} \right)  \overline{\rho},
\end{equation}

\noindent as the equivalent of the classical tidal density criterion, as a first generic approximation under modified gravity.
Since the spatial extent of wide binaries will always be much smaller that their Galactocentric radii, equation (5)
shows that under modified gravity, to first order, wide binaries will be much more robust to tides than 
under Newtonian gravity. 
In the following section we apply the test we have identified to two recent catalogues of wide binaries  
which became available over the previous year, the SLoWPoKES catalogue of SDSS wide binaries by \citep{11}
and the {\it Hipparcos} satellite wide binaries catalogue of \citep{29}.

\section{Observed wide binary Samples}

\subsection{The Hipparcos wide binaries}

\begin{figure}[!t]
\includegraphics[angle=0,scale=0.44]{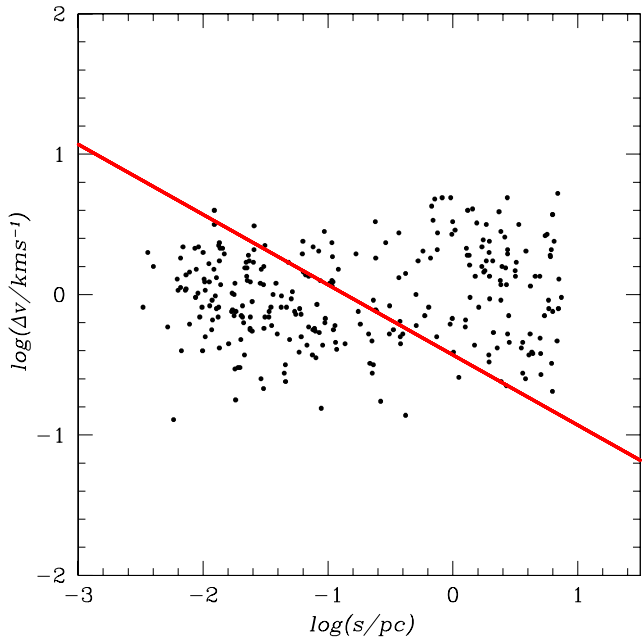}
\caption{The figure shows projected relative velocities and separations for each pair of
wide binaries from the \citep{29} {\it Hipparcos} catalogue having a probability of being 
the result of chance alignment $<0.1$. The average value for the signal to noise ratio for the sample shown 
is 1.7. The upper limit shows the flat trend expected from modified gravity theories, at odds
with Kepler's third law, shown by the $s^{-1/2}$ solid line.}
\end{figure}

The \citep{29} catalogue of very wide binaries was constructed through a full Bayesian analysis of the 
combined {\it Hipparcos} database, the new reduction of the {\it Hipparcos} catalogue, \citep{32}, the Tycho-2 
catalogue, \citep{15} and the {\it Tycho} double star catalogue, \citep{12} mostly, amongst others.
There, probable wide binaries are identified by assigning a probability above chance alignment to the stars analysed by carefully 
comparing to the underlying background (and its variations) in a 5 dimensional parameter space of proper motions and spatial 
positions. The authors have taken care to account for the distortions introduced by the spherical projection on the relative 
proper motion measurements, $\Delta \mu$. When angular separations cease to be small, small relative physical velocities between 
an associated pair of stars might result in large values of $\Delta \mu$. A correction of this effect is introduced, to keep 
$\Delta \mu$ values comparable across the whole binary separation range studied.

We have taken this catalogue and kept only binaries with a probability of non-chance alignment greater than $0.9$. 
The wide binary search criteria used by the authors requires that the proposed binary should have no near neighbours; 
the projected separation between the two components is thus always many times smaller than the typical interstellar 
separation, see \citep{29}. We use the reported distances to the primaries, where errors are smallest, 
to calculate projected $\Delta V$ and projected $s$ from the measured $\Delta \mu$ and $\Delta \theta$ values reported 
by \citep{29}. Although the use of {\it Hipparcos} measurements guarantees the best available quality in 
the data, we have also further pruned the catalogue to remove all binaries for which the final signal to noise ratio in 
the relative velocities on the plane of the sky was lower than $0.3$.

We plot in figure (1) a sample of 280 binaries constructed as described above, having distances to the Sun within 
$6<d<100$ in pc. The slanted line gives the Newtonian prediction of eq.(1) to the upper limit expected on the relative 
velocities shown, which appears in conflict with it, as they are defined by a neat horizontal upper limit, as 
generically predicted by modified gravity theories, eq.(2). Figure (1) could then be a first direct evidence of the 
breakdown of classical gravity theories in the low acceleration regime of $a<a_{0}$.

The average signal to noise ratio for the data in figure (2) is 1.7, with an average error on $\Delta V$ of 0.83 $km/s$, 
which considering a $2 \sigma$ factor from the top of the distribution to the real underlying upper limit for the sample, 
results in 3 $km/s$ as our estimate of the actual physical upper limit in $\Delta V$. Comparing with eq.(2), the factor 
accounting for non-circular orbits in modified gravity comes to 4.5. That this factor is significantly larger than the 
$\sqrt2$ of Newtonian gravity is to be expected, as objects are much more tightly bound in MOND-type schemes.

\begin{figure}[!t]
\includegraphics[angle=0,scale=0.44]{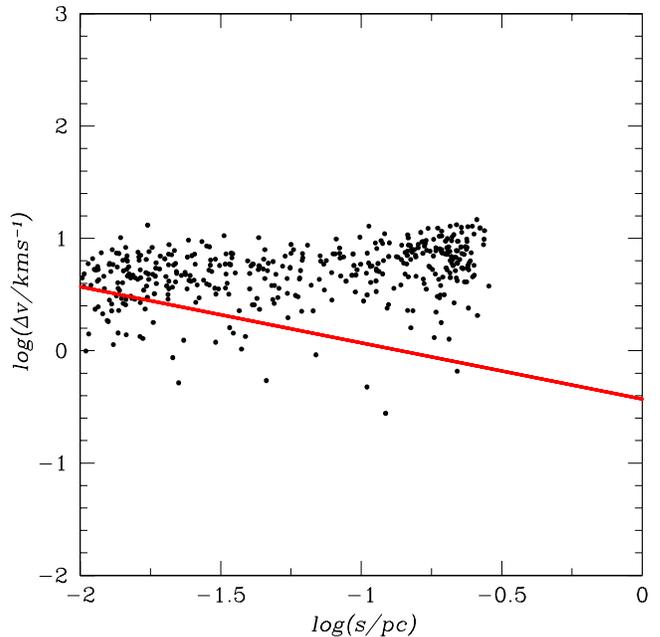}
\caption{The figure shows projected relative velocities and separations for each pair of
wide binaries from the \citep{11} SDSS catalogue within the distance range
($225<d/pc<338$). The average value for the signal to noise ratio for the sample shown 
is 0.5. The upper limit shows the flat trend expected from modified gravity theories, at odds
with Kepler's third law, shown by the $s^{-1/2}$ solid line.}
\end{figure}

\begin{figure}[!t]
\includegraphics[angle=0,scale=0.44]{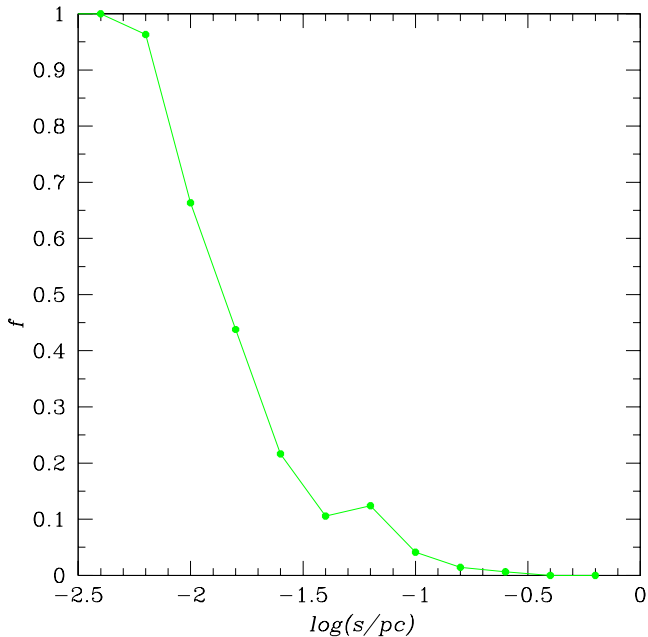}
\caption{The figure shows the number of systems with $\Delta V$ values below the Newtonian upper limit prediction of eq. (1),
as a fraction of the total per bin, for the SDSS data of \citep{11}, as a function of projected binary separations.}
\end{figure}

\subsection{The SDSS wide binaries}

The Sloan low mass wide pairs catalogue (SLoWPoKES) of \citep{11} contains a little over 1,200 wide 
binaries with relative proper motions for each pair, distances and angular separations. Also, extreme care was 
taken to include only physical binaries, with a full galactic population model used to exclude chance alignment 
stars using galactic coordinates and galactic velocities, resulting in an estimate of fewer than 2\% of false 
positives. As with the {\it Hipparcos} sample, this last requirement yields only isolated binaries with no 
neighbours within many times the internal binary separation. We have also excluded all systems with white dwarfs 
or subdwarf primaries, where distance calibrations are somewhat uncertain. As was also done for the {\it Hipparcos}
sample, all triple systems reported in the catalogue were completely excluded from the analysis.

Given the large range of distances to the SDSS binaries ($46<d/pc<992$), we select only 1/3 of the sample lying within the narrow 
distance range ($225<d/pc<338$), which forms the most homogeneous set in terms of the errors in $\Delta V$, excluding data with 
large errors at large distances. Again, we use the reported distances to the primaries, where errors are smallest, to calculate 
projected $\Delta V$ and projected $s$ from the measured $\Delta \mu$, $\Delta \theta$ and $d$ values reported by \citep{11}, 
to plot figure (2). The figure shows 417 binaries with average signal to noise ratio and average errors on $\Delta V$ of 
0.5 and 11.3 $km/s$, respectively.

The slanted solid line gives the Newtonian prediction of eq.(1). It is clear that the upper envelope of the distribution of 
$\Delta V$ measurements from the catalogue does not comply with Kepler's third law. As was the case with the {\it Hipparcos} 
sample, the upper envelope of the distribution of observed measurements describes a flat line, as expected under modified gravity 
schemes.

Figure (3) is a plot of the number of SLoWPoKES systems for the full distance range, with $\Delta V$ values below the Newtonian 
prediction of eq.(1), and hence consistent with it, as a fraction of the total per bin, as a function of projected binary 
separations. We see this fraction starts off being consistent with 1, but begins to decrease on approaching $log(s/pc) \approx 
-2$, after which point it rapidly drops, to end up consistent with 0 on reaching separations of around $log(s/pc) \approx -1$. 
Again, the result matches the qualitative generic expectations of modified gravity schemes, but would call for further 
explanations under classical gravity.

The average signal to noise values in $\Delta V$ for the full distance range of the \citep{11} catalogue is
0.48. The average error on $\Delta V$ for the full SDSS sample is 12 $km/s$, which considering a $2 \sigma$ factor from the top 
of the complete distribution to the real underlying upper limit, results in the same $3 km/s$ as obtained for the \citep{29} 
{\it Hipparcos} catalogue.

\begin{figure}[!t]
\includegraphics[angle=0,scale=0.44]{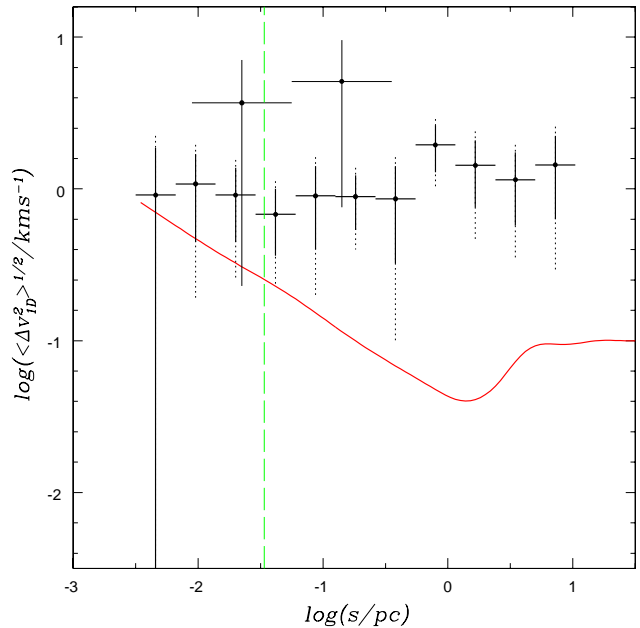}
\caption{The solid curve gives the RMS values for one dimensional projected relative velocities as a function of
projected separations, for the detailed dynamical modelling of large populations of wide binaries evolving in
the Galactic environment, taken from \citep{16}. The same quantity for the data from the catalogues analysed is
given by the points with error bars; those with narrow $log(s)$ intervals being from the {\it Hipparcos} sample of \citep{29},
and those two with wide $log(s)$ intervals coming from the SDSS sample of \citep{11}.}
\end{figure}

We end this section with figure (4), where we calculate the RMS value of the one-dimensional relative velocity difference for 
both of the samples discussed, after binning the data into constant logarithmic intervals in $s$. This quantity is given by the 
points, where the error bars simply show the propagation of the errors on $\Delta \mu$ and $d$, reported by the authors of the 
catalogues. We construct $\Delta V_{1D}$ by considering only one coordinate of the two available from the relative motion on the 
plane of the sky. Thus, each binary can furnish two $\Delta V_{1D}$ measurements, which statistically should not introduce any 
bias. Indeed, using only  $\Delta \mu_{l}$ or only $\Delta \mu_{b}$ or both for each binary, yields the same mean values for the 
points shown. The small solid error bars result from considering an enlarged sample where each binary contributes two $\Delta V_
{1D}$ measurements, while the larger dotted ones come from considering each binary only once, and do not change if we consider 
only $\Delta \mu_{l}$ or only $\Delta \mu_{b}$. The series of small $log(s)$ interval data are for the {\it Hipparcos} catalogue 
of \citep{29}, while the two broader crosses show results for the \citep{11} SDSS sample of figure (2). 
For this last case, the much larger intrinsic errors mean that to compensate through weight of numbers the low signal to noise 
ratio of this catalogue, imposes the loss of separation resolution through the use of only two bins, with vertical error bars 
which are only relevant if the sample is doubled, as described above.

The solid curve is the Newtonian prediction of the full Galactic evolutionary model of \citep{16} for a randomly
oriented population of wide binaries with a realistic distribution of eccentricities, both bound and in the process of 
dissolving. Note that the results of this simulation deviate from Kepler's law for s larger than the Newtonian Jacobi radius of 
the problem of $1.7 pc$, whereas the deviation shown by 
the samples of binaries studied also occur at much smaller separations (see below).  Even considering the large error bars, where 
each binary contributes only one $\Delta V_{1D}$ value, we see eight points lying beyond 1$\sigma$, making the probability of 
consistency between this prediction and the observations of less than $(0.272)^{8}$=$3\times 10^{-5}$. The only point where this 
model is slightly at odds with the selection of the {\it Hipparcos} sample of \citep{29}, is that \citep{16}  
assume $1 M_{\odot}$ stars for their binaries, while the typical mass of the stars in the binaries we examine is closer to $0.5 
M_{\odot}$. This detail would only shift the Newtonian prediction a factor of $2^{1/2}$ further away from the measurements. We 
obtain an RMS value for $\Delta V_{1D}$ compatible with a horizontal line at 1 $km/s$, in qualitative agreement with expectations 
from modified gravity schemes. The vertical dashed line marks the $a=a_{0}$ threshold; we clearly see the data departing from the 
Newtonian prediction outwards of this line, and not before.

Under Newtonian Gravity one would need to look for an alternative dynamical dissociation and heating mechanism for the binaries 
we analyse which might result in relative velocities an order of magnitude above the results obtained by \citep{16}.
The {\it Hipparcos} catalogue has been closely studied for over a decade, and not only the values reported, but also the
uncertainties in them are now well established. It is highly unlikely that these confidence intervals might have been systematically
underestimated by the community by a factor of 3, as would be required for the data in fig.(4) to be consistent with the Newtonian
prediction of \citep{16}.

The two SDSS points are clearly consistent with the {\it Hipparcos} measurements, but given the much larger error bars, 
they are also marginally consistent with the Newtonian prediction at a 2$\sigma$ level. A fuller discussion of the 
robustness of our results to the various sample selection effects and errors involved appears in the Appendix.

The consistency of the results obtained from both catalogues provides a check of the physical reality of the trends presented.
The two completely independent, very carefully constructed catalogues, each using different sets of selection criteria, each 
perhaps subject to its own independent systematics, are consistent with the same result, a constant horizontal upper envelope for 
the distribution of relative velocities on the plane of the sky at an intrinsic value of 3 $km/s \pm$1 $km/s$, extending over 3 
orders of magnitude in $s$, with a constant RMS $\Delta V_{1D}$ value consistent with 1 $km/s \pm$0.5 $km/s$. This supports the 
interpretation of the effect detected as the generic prediction of modified gravity theories.

Given the relevance of the subject matter discussed, it would be highly desirable to obtain an independent confirmation, ideally 
using a purposely designed sample including cuts at a variety of stellar masses, in order to test the scaling of the fourth power 
of the dynamical velocities with mass expected under modified gravity schemes. To include also radial velocity measurements would 
require sampling over many epochs, as the only way to allow for the effects of nearby companions in these frequently hierarchical 
systems.

\section{Conclusions}

We have identified a critical test in the classical gravity/ modified gravity debate, using the relative velocities of wide 
binaries with separations in excess of 7000 AU, as these occupy the $a<a_{0}$ regime characteristic of modified gravity models.
We present a first application of this critical test using the best currently available data; a large sample of wide binaries 
from the SDSS with low signal to noise on the relative velocities, and a smaller {\it Hipparcos} satellite sample with signal to 
noise $\sim 2$ on the relative velocities of the binaries sampled.

Results show constant  relative RMS velocities for the binary stars in question, irrespective of their separation, in the $a<a_
{0}$ regime sampled. This is quantitatively inconsistent with detailed predictions of Newtonian dynamical models for large 
populations of binaries evolving in the local galactic environment.

Our results are qualitatively in accordance with generic modified gravity models constructed to explain galactic dynamics in the 
absence of dark matter, where one expects constant relative velocities for binary stars, irrespective of their separation, in the 
$a<a_{0}$ regime sampled.

\section{Acknowledgements}

The authors wish to thank Arcadio Poveda and Pavel Kroupa for insightful discussions on the subject treated here. Alejandra 
Jimenez acknowledges financial support from a CONACYT scholarship. Xavier Hernandez acknowledges financial support from UNAM-
DGAPA grant IN103011.

\appendix

\section{Calculation of confidence intervals}

We begin this section with a discussion of various analyses performed to test for the possibility that 
our results could have been driven by potential systematics and selection effects in the catalogues.

We first test for the option that the results of figures (1) and (2) were distorted from the
Newtonian prediction by errors which correlate with the separation of the binaries, increasing as the separation
increases, to yield the trends obtained. For the {\it Hipparcos} sample we ranked the binaries by separation, $s$, and calculated
the average errors on the resulting $\Delta V$ in each the tightest third, middle third and widest third of the binaries, 
yielding values 0.8 (0.4), 0.7 (0.11) and 1.0 (0.4) respectively, in $km/s$. The numbers in parenthesis giving the 
dispersion of the distributions of errors in each of the three thirds of the sample. It can be seen that there is no 
increase either in the average values of the errors in $\Delta V$, or in the width of the distributions of errors, 
with increasing binary separations. For the SDSS sample the average errors in $\Delta V$ show only a very slight 
increase of a factor of 1.3 over the entire range probed. Thus, for both samples, the trends of figures (1), (2) and 
(4) can not be explained as arising from the Newtonian prediction and an increase in the errors with binary separation.
In essence, the data presented are inconsistent with the Newtonian prediction not because of differences in the
details of the trends, but because the former presents multiple real detections at a level of $3 km/s \pm 1km/s$, 
while the latter is smaller by an order of magnitude. The resulting RMS values for the various data points are of 
$\approx 1 km/s \pm 0.5 km/s$, while the Newtonian prediction lies below the detection by about a factor of 10. 

We next check against systematics in the catalogues, which would preferentially arise in the lower signal to noise points, or 
alternatively, in the most distant ones. For the {\it Hipparcos} sample, we repeat the experiment including only points with a
signal to noise ratio on $\Delta V >1$, and then including only the binaries nearest to earth with $d< 50 pc$. These two sub-
samples have 186 and 130 binaries, with average signal to noise ratios on $\Delta V$ above the value of 1.7 of figure (1), of
2 and 2.2 respectively, and yield results indistinguishable from figs. (1) and (4). Also, we checked explicitly the results of 
the SDSS sample for any systematics with distance to the binaries analysed, and found the flat upper envelope to be robust to 
the choice of distance range taken.

Although no specific cut in resulting $\Delta V$ or even on measured $\Delta \mu$
was built into the {\it Hipparcos} catalogue, the SDSS one
includes the selection cut that the signal to noise ratio on $\Delta \mu$
should be $<2$. We repeat the SDSS analysis tracing the 15\% of the points closest to
this cut, and find that these, the ones which just made the cut, do not define the
flat upper envelope on $\Delta V$. For both the {\it Hipparcos} and the SDSS
samples, any observational bias/truncation would appear in ($\Delta \mu$, $\Delta \theta$) 
space, not in the ($\Delta V$, $s$) one.

We end with a detailed description of the calculation of the points in figure (4) and their error bars, directly from the data 
published in the catalogues used. Whenever $z=f(x)$, the error in $z$ is given by $\delta z=|\frac{df}{dx}|\delta x$, 
where $\delta x$ is the error in $x$. If $z$ is a function of several variables 
 $z=f(x_{1},x_{2},...)$, the absolute error, which we take, is $\delta z=\sum_{i}|\frac{\partial f}{\partial x_{i}}|_{x_{i}}\delta x_{i}$ which 
is always greater than standard deviation $\sigma_{z}= \sqrt{\sum_{i}(\frac{\partial f}{\partial x_{i}}\delta x_{i})^{2}}$.

Following these rules we estimate the absolute error in each $\Delta V$ measurement, and in the RMS velocity for the samples 
studied. For the \textit{Hipparcos} sample we have taken columns 11,12,13 and 14 of the online catalogue of \citep{29}, 
which contains the values of the relative motions of each binary and their errors, as well as columns 15 and 16, which 
contain the distance to the binary, $d$, and its corresponding error, $\delta d$.

For all calculations, we considered the distance to the system as the distance to the primary. For each binary we calculated the 
projected relative velocity $\Delta v$ in $km/s$  and its error as:

\begin{equation}
\Delta v=4.74\times 10^{-3}\Delta\mu d, 
\end{equation}

\begin{equation}
 \delta(\Delta v)=4.74\times 10^{-3}(\delta(\Delta\mu)d + \Delta\mu\delta d),
\end{equation}

\noindent hence, on $(\Delta v)^{2}$the error is $\delta(\Delta v)^{2}=2\Delta v \delta(\Delta v)$.
The RMS velocity is now,

\begin{equation}
 \Delta v_{RMS}=\sqrt{\frac{1}{n}\sum_{i=1}^{n}(\Delta v)^{2}}=\sqrt{<(\Delta v)^{2}>}.
\end{equation}

We then binned the data into constant logarithmic intervals in $s$ with a width of 0.32 and calculated $\Delta v_{RMS}$ for each 
bin, $n$ the number of binary systems that fall into each bin. The error in  $\Delta v_{RMS}$ for each bin is now:

\begin{equation}
\delta(\Delta v_{RMS})=\frac{1}{2}\frac{<\delta(\Delta v)^{2}>}{\sqrt{<(\Delta v)^{2}>}}.
\end{equation}

To compare with the prediction of the RMS values for the one dimensional projected relative velocities of the evolutionary 
model of \citep{16} we construct $\Delta v_{RMS}$ by considering only one coordinate. For the case of the SDSS 
sample of \citep{11}, we have calculated the relative proper motion for each coordinate $\Delta\mu_{\alpha}=\mu_
{\alpha_{A}}-\mu_{\alpha_{B}}$ and $\Delta\mu_{\delta}=\mu_{\delta_{A}}-\mu_{\delta_{B}}$ taking columns 18 to 25 of their online
catalogue, which contain the proper motions and their errors in equatorial coordinates for each component of the binary system, to 
estimate the projected relative velocity we have used column 26 containing the distance to the primary, in this case we consider 
an error of $15\%$ in the distance and we have followed the same path as in the previous case for calculating the RMS relative 
velocities and their errors.


\end{document}